\documentclass[useAMS,usenatbib]{mn2e}

\usepackage{graphicx}
\usepackage{color}

\begin{document}

\newcommand{\rat}[0]{\lbrack O\,{\sc III}\rbrack/H\,$\beta$}


\title[Interpreting high \lbrack O\,{\sc III}\rbrack /H\,$\beta$ ratios]{Interpreting high  \lbrack O\,{\sc III}\rbrack /H$\bbeta$ ratios with maturing starbursts}
\author[E.~R.~Stanway et al.]{Elizabeth R.~Stanway$^{1}$\thanks{E-mail:
e.r.stanway@warwick.ac.uk}, John J.~Eldridge$^{2}$, Stephanie M.~L.~Greis$^{1}$,\newauthor Luke J.~M.~Davies$^{3}$, Stephen M.~Wilkins$^{4}$, Malcolm N.~Bremer$^{5}$\\
$^{1}$Department of Physics, University of Warwick, Gibbet Hill Road, Coventry, CV4 7AL, UK\\
$^{2}$Department of Physics, University of Auckland, Private Bag 92019, Auckland, New Zealand\\
$^{3}$ICRAR, The University of Western Australia, 35 Stirling Highway, Crawley, WA 6009, Australia\\
$^{4}$Astronomy Centre, Department of Physics and Astronomy, University of Sussex, Brighton, BN1 9QH, U.K\\
$^{5}$H.~H.~Wills Physics Laboratory, University of Bristol, Tyndall Avenue, Bristol, BS8 1TL, UK}

\date{Accepted 2014 August 15. Received 2014 August 05; in original form 2014 May 28}

\pagerange{\pageref{firstpage}--\pageref{lastpage}} \pubyear{2014}

\maketitle

\label{firstpage}

\begin{abstract}
Star forming galaxies at high redshift show ubiquitously high ionization parameters, as measured by the ratio of optical emission lines. We demonstrate that local ($z<0.2$) sources selected as Lyman break analogues also manifest high line ratios with a typical \lbrack O\,{\sc III}\rbrack /H\,$\beta=3.36^{+0.14}_{-0.04}$ - comparable to all but the highest ratios seen in star forming galaxies at $z\sim2-4$. We argue that the stellar population synthesis code BPASS can explain the high ionization parameters required through the ageing of rapidly formed star populations, without invoking any AGN contribution. Binary stellar evolution pathways prolong the age interval over which a starburst is likely to show elevated line ratios, relative to those predicted by single stellar evolution codes. As a result, model galaxies at near-Solar metallicities and with ages of up to $\sim$100\,Myr after a starburst typically have a line ratio \lbrack O\,{\sc III}\rbrack /H\,$\beta$$\sim$3, consistent with those seen in Lyman break galaxies and local sources with similar star formation densities. This emphasises the importance of including binary evolution pathways when simulating the nebular line emission of young {\color{black}or bursty} stellar populations.
\end{abstract}


\section{Introduction}\label{sec:intro}

Understanding the sites of star formation in the distant Universe is key to developing our picture of the early stages of galaxy evolution. The low mass, low metallicity, intensely star forming galaxies observed in deep field surveys are the building blocks which form the more massive systems we currently inhabit and observe evolve. They are also the most likely source of the energetic photons that ionized the intergalactic medium (IGM) at early times \citep[e.g.][]{2010MNRAS.409..855B}, creating the conditions which persist to the current day. The source and spectrum of those ionizing photons are key parameters in cosmological simulations, affecting the process by which small regions of ionized Hydrogen surrounding the first galaxies grow and eventually overlap.

However, spectroscopy of such distant sources pushes the technical limits of existing spectrographs, and is often impossible. Only the most highly lensed galaxies, or extreme examples such as submillimetre galaxies or quasar hosts, are sufficiently luminous to measure optical emission lines at $z>5$, where the rest-UV and optical wavelength ranges have been shifted into the near-infrared. Nonetheless, the fitting of spectral energy distributions (SEDs) across a broad redshift range ($z\sim3-8$) has strongly suggested that the majority of high redshift star forming galaxies may show strong optical line emission, which contributes significantly to their observed flux at 3-5\,microns \citep{2014A&A...563A..81D,2013ApJ...763..129S,2012ApJ...755..148G}. Local galaxies selected to have similar ultraviolet emission densities are also confirmed to have very prominent emission lines \citep{2005ApJ...619L..35H,2014MNRAS.439.2474S}. 

At slightly lower redshifts, $z\sim2-4$, direct measurements of the rest-optical emission spectrum become possible.  Work by \citet{2014arXiv1401.5490H} compiled $K$-band spectroscopy on a sample of 67 $z\sim3.5$ galaxies, determining that high \lbrack O\,{\sc III} 5007\AA\rbrack/H\,$\beta$ ratios are ubiquitous in the high redshift population. More recently still, \citet{2014arXiv1405.5473S} compiled a large sample of 179 $2.0<z<2.6$ objects with near-infrared spectroscopy and confirmed that these high redshift, star forming galaxies occupy a distinct locus with higher line ratios than seen in typical local galaxies.  However, efforts to model these high ratios are problematic, requiring ionization parameters orders of magnitudes higher than those typically seen in local galaxies \citep{2000ApJ...542..224D,2013ApJ...774..100K,2001ApJ...556..121K} or invoking an otherwise unrealistic high oxygen abundance \citep{2014A&A...564A..19C}. 

In this letter we present measurements of the line ratios determined in a sample of local galaxies which are selected to match the distant population in photometric properties (section \ref{sec:data}), and consider possible interpretations in the light of models from the Binary Population and Spectral Synthesis (BPASS) models (section \ref{sec:bpass}). In section \ref{sec:discussion} we discuss implications for our understanding of stellar populations in the distant Universe, before presenting our conclusions.

\section{Observed optical emission line ratios}\label{sec:data}

 \subsection{Evidence for high line ratios at $z>1$}

The ratios of optical spectral lines, arising from nebular emission, are dependent on the combination of ionizing ultraviolet flux incident on the ISM and its density. Amongst them, one of the most useful in the distant Universe is the ratio of the [O\,{\sc III}] 5007\AA\ line to the H\,$\beta$ 4861\AA\ Balmer line. The small wavelength interval between these lines minimises the impact of dust absorption uncertainties on their ratio, and they are more accessible at high redshifts ($z>1$) than the redder H$\alpha$ line region. Samples of distant star-forming galaxies with rest-frame optical spectroscopy have remained small nethertheless, largely due to the challenging nature of the observations which have required observations of individual targets from the ground or low resolution grism spectra from the {\em Hubble Space Telescope} \citep[e.g.][]{2012AJ....144...28X}.  Lensed targets have been the most studied, for example in \citet{2009ApJ...701...52H} who used \rat\ and other line ratios to measure high ionization parameters in three lensed galaxies at $z\sim2-2.5$. 

\begin{figure*}
 \includegraphics[width=\columnwidth]{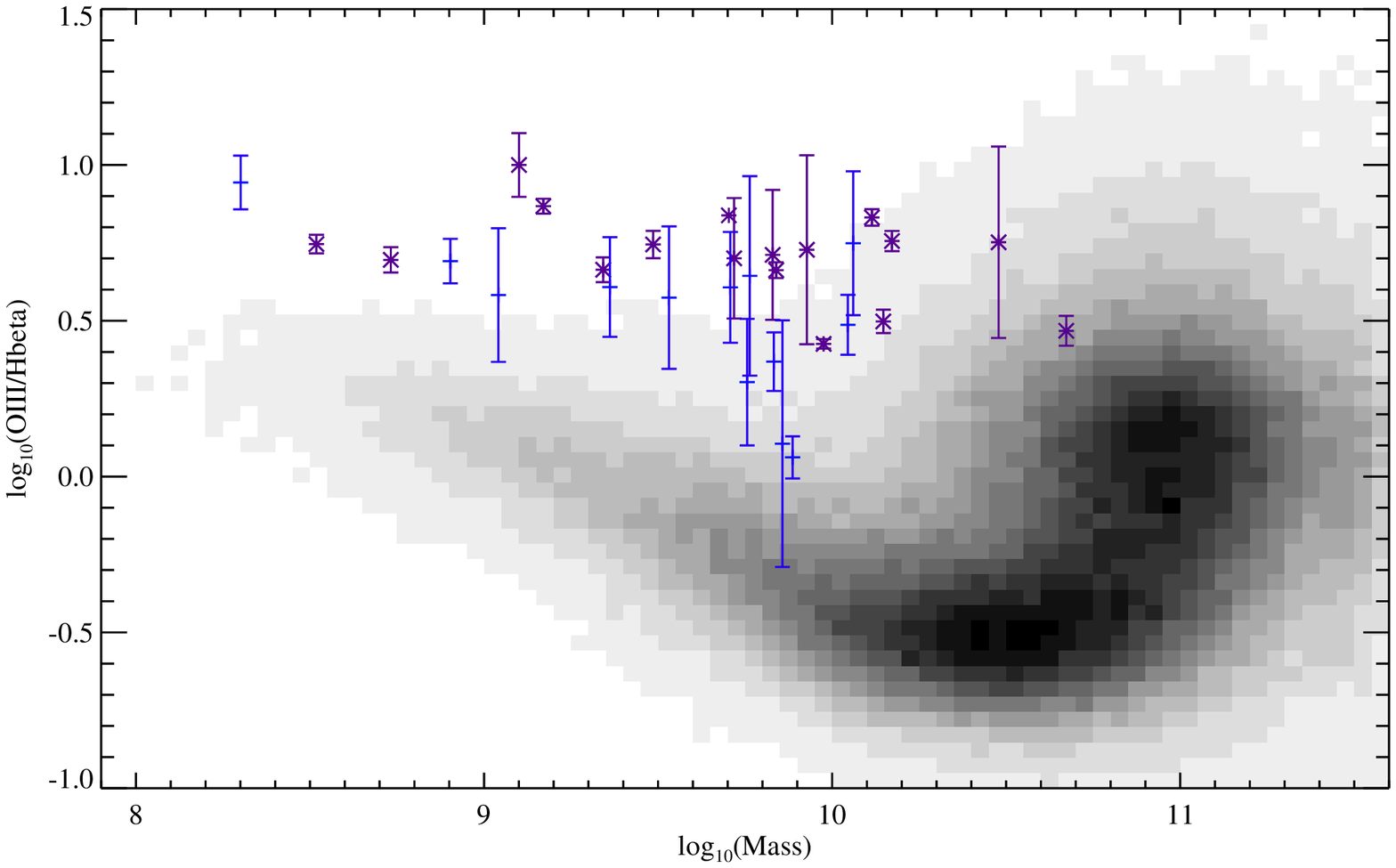}
 \includegraphics[width=\columnwidth]{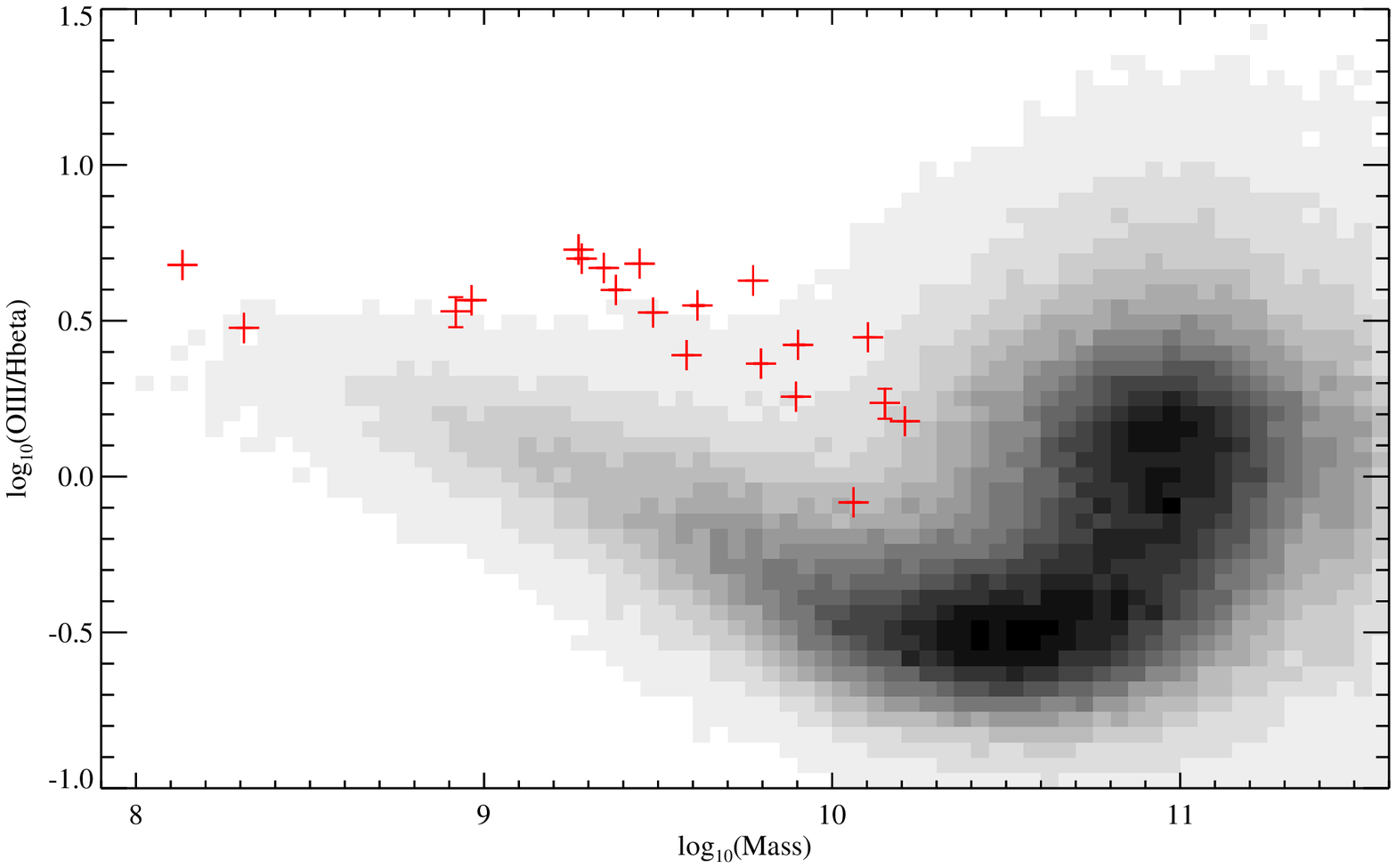}
 \caption{The distribution of  OIII   /H$_\beta$ ratios seen in galaxies observed by the SDSS (Brinchmann et al 2004, greyscale). Overplotted on the left are the ratios seen at $z\sim2-4$ by Holden et al (2014, asterisks), and Schenker et al (2013, crosses). The ratios measured in the Lyman break analogue sample of \citet{2014MNRAS.439.2474S}, and presented here for the first time, are shown on the right. The measurement error on the spectra of these bright sources is smaller than the plotting symbols. The mass uncertainty in any SED fitting analysis is at least $\sim$0.3\,dex. At both high and low redshift, UV-selected galaxies exceed the typical line ratios for their mass. Note that the high mass, high line ratio sources in the local SDSS sample are typically AGN dominated.\label{fig:ratios}}
 \end{figure*}

As multi-object near-infrared spectrographs become available, analyses of larger samples are possible. \citet{2013ApJ...769....3N} found high ionization parameters in $z=2.2$ starforming galaxies selected as Lyman alpha emitters, while \citet{2014ApJ...785..153M} measured a \rat\ ratio of 4.12$\pm$0.01 in a composite spectrum of 24 $z\sim2$ galaxies, exceeding the predictions of photoionization models. Recent work by \citet{2014arXiv1401.5490H} compiled $K$-band spectroscopy on a sample of 67 $z\sim3$ ultraviolet-selected `Lyman break' galaxies \citep[including 20 observed by][]{2013ApJ...777...67S}, determining a median of \rat $=4.8^{+0.8}_{-1.7}$ for the sample. In figure \ref{fig:ratios} we replot the high \rat\ line ratios measured by \citeauthor{2014arXiv1401.5490H} as a function of mass, together with the distribution of ratios observed in low redshift galaxies \citep[][]{2004MNRAS.351.1151B}. Distant galaxies lie well above the line ratios seen in local star forming galaxies at the same mass. The hard ionizing radiation field required to explain these observations is very difficult to reproduce with normal stellar populations \citep{2013ApJ...774..100K,2008MNRAS.385..769B} at the metallicities seen at $z\sim2-5$ \citep[typically 0.1-1.5\,Z$_\odot$, e.g.][]{2011MNRAS.413..643R,2009ApJ...701...52H,2006ApJ...644..813E,2001ApJ...554..981P}.

 \subsection{Analogues in the local universe}

  While the vast majority of galaxies in the local Universe differ in size, star formation rate and character from those at high redshift, it is nonetheless possible to identify local sources which appear to match the distant population in their continuum properties. In our recent paper \citep{2014MNRAS.439.2474S}, we identified a pilot sample of such analogues, lying at $z\sim0.05-0.25$. Galaxies were identified based on their ultraviolet luminosity and colour, and to ensure that the resulting sample matched the high specific star formation densities observed at $z\sim5$. Given that high line ratios appear ubiquitous at $z>1$, we would expect these to share that characteristic.

Our sample were also selected to have spectroscopy from the Sloan Digital Sky Survey (SDSS), precluding the presence of strong AGN activity and confirming their redshift. In figure \ref{fig:ratios} we present the \rat\ line ratios measured in our local analogue sample, as a function of stellar mass. Line ratios are measured at high signal to noise in SDSS spectroscopy for these relatively bright sources - we use line fluxes calculated by the SDSS pipeline, but have checked that these agree to within 0.02 dex with the independent MPA-JPU DR7 analysis\footnote{see www.mpa-garching.mpg.de/SDSS/DR7 \citep{2004MNRAS.351.1151B}}. 
Stellar masses are also taken from the MPA-JPU database. These are derived from simultaneous fitting of age, dust, star formation history and stellar mass to the galaxies' spectral energy distribution. Thus they are uncertain at the $\sim0.3$\,dex level, due to reliance on the synthesis model templates used for fitting and degeneracies in the resulting colours. The uncertainties in mass fitting of these sources will be explored in more detail in Greis et al (2014, in prep).

As figure \ref{fig:ratios} demonstrates, our analogue sample also lies well above the norm for line ratios in star forming galaxies of the same mass in the local Universe \citep{2004MNRAS.351.1151B}, despite being drawn from the same underlying data set. While the analogue sample does not match the most extreme high redshift examples, the ultraviolet continuum selection of our Lyman break analogue sample appears to identify galaxies with very comparable spectral properties to those seen in the distant Universe.  We find that the low redshift analogue sample presented here has a median of \rat $=3.36^{+0.14}_{-0.04}$, where the quoted uncertainty gives the inter-quartile range, consistent with that quoted by \citet{2014arXiv1401.5490H} for their $z\sim2-3$ sample.

\section{Modelling high ratios with BPASS}\label{sec:bpass}

 \subsection{BPASS}\label{sec:bpass2}
  The Binary Population and Spectral Synthesis (BPASS) models\footnote{see http://bpass.org.uk/} are a set of galaxy population synthesis models which were developed to address the effects of massive stars on the spectral energy distributions of galaxies \citep{2012MNRAS.419..479E,2009MNRAS.400.1019E,2008MNRAS.384.1109E}.  Given a young stellar population, for example in the aftermath of a major star formation episode, the optical spectrum of galaxies is dominated by hot and massive stars which have not yet reached the ends of their lifespan. However, as the population ages, {\color{black}the population averaged colour, temperature and SED are all strongly influenced by the evolutionary state of the remaining massive stars}. The processes of angular momentum transfer, mass loss or mass gain due to a binary companion all modify this evolutionary state, allowing evolved secondary stars to extend the highly-luminous phase, and boosting the population of rapidly rotating, hydrogen-depleted, Wolf-Rayet stars.

The BPASS code tracks the evolution of stellar populations, sampled from an initial mass function and range of binary system properties, and creates a composite stellar spectrum at a given age. Binary evolution is treated explicitly for initial stellar masses $\gid 5$\,M$_\odot$, and empirical terms are included for the evolution of rotationally-mixed, quasi-homogeneous stars. For comparison, an equivalent population of single stars is also permitted to evolve. The radiative transfer of the stellar emission from both populations, through a dust and gas screen within the source galaxy, is then modelled using the radiative tranfer code {\sc CLOUDY} \citep{1998PASP..110..761F} to assess the contribution of nebular continuum and line emission.  {\color{black}The ionization parameter of the local radiation field is
defined as a ratio of the number of ionizing photons to the local gas density. It is thus not a directly tunable parameter, but rather constructed through the combination of appropriate stellar atmosphere models and a choice of gas distribution}. The assumed total Hydrogen gas density of our baseline model set is $10^2$\,cm$^{-3}$, distributed in a sphere around the stellar population. This is a fairly typical gas density for extragalactic star forming H\,{\sc II} regions, although we note that these range over several orders of magnitude in density \citep[see e.g.][]{2009A&A...507.1327H}. We explore effects of varying the gas density relative to our baseline models in section \ref{sec:gas}. The evolution of an instanteous, rapid burst of star formation and of a continous moderate (1\,M$_\odot$\,yr$^{-1}$) star formation rate are modelled separately. In this letter we use nebular emission line flux predictions determined as part of the current (v1.0) BPASS model data release.

 \begin{figure*}
 \includegraphics[width=\columnwidth]{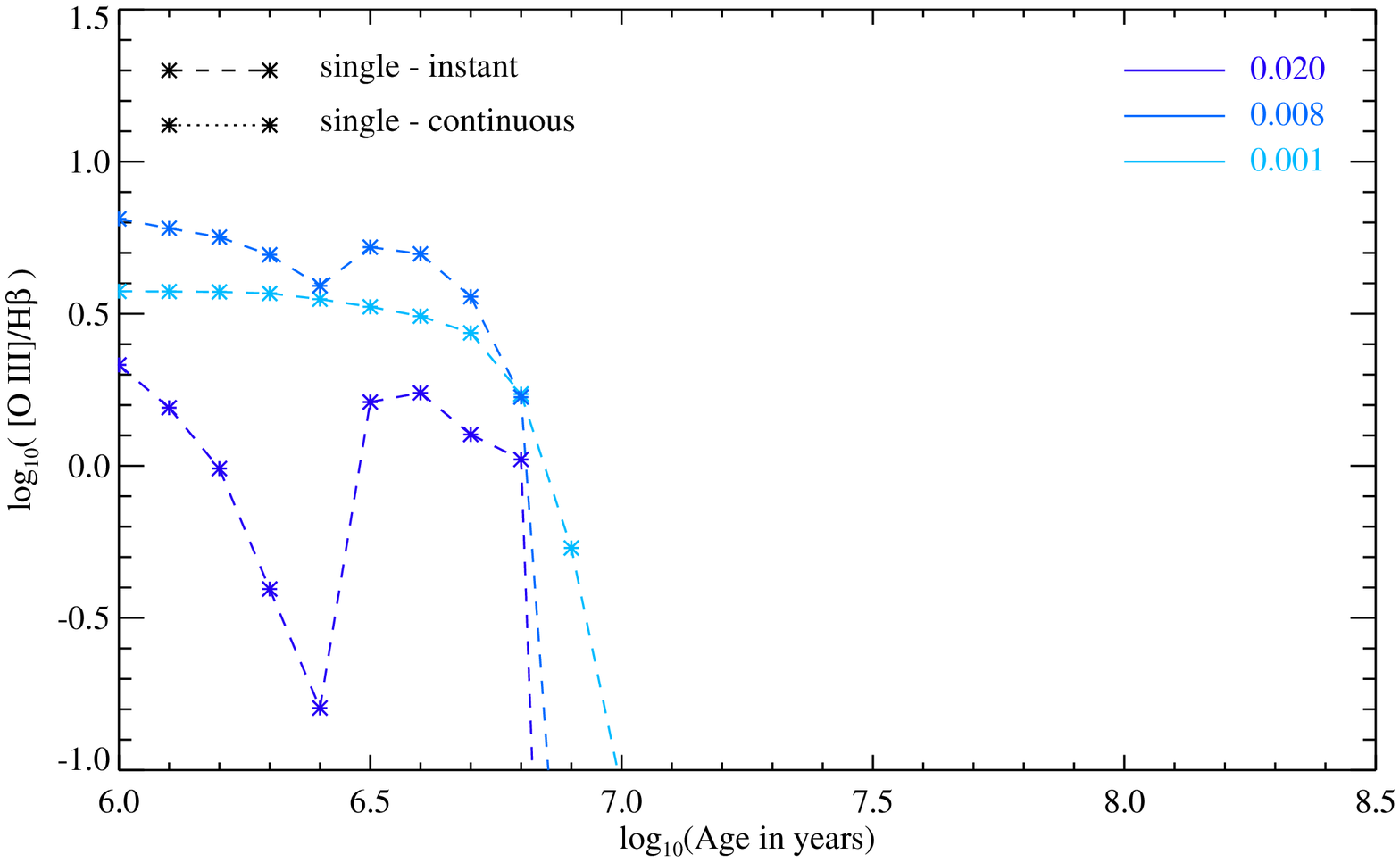}
 \includegraphics[width=\columnwidth]{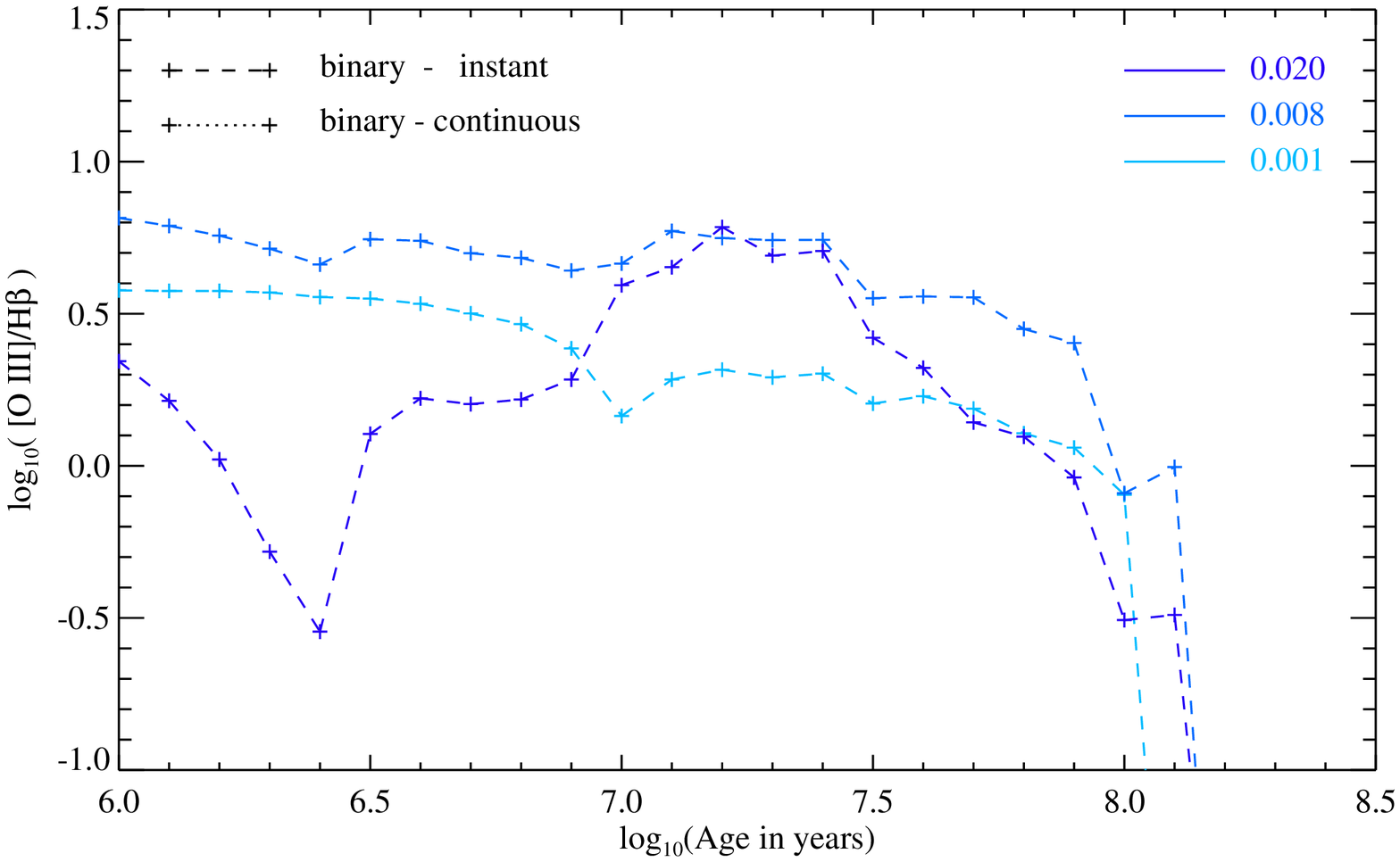}
 \includegraphics[width=\columnwidth]{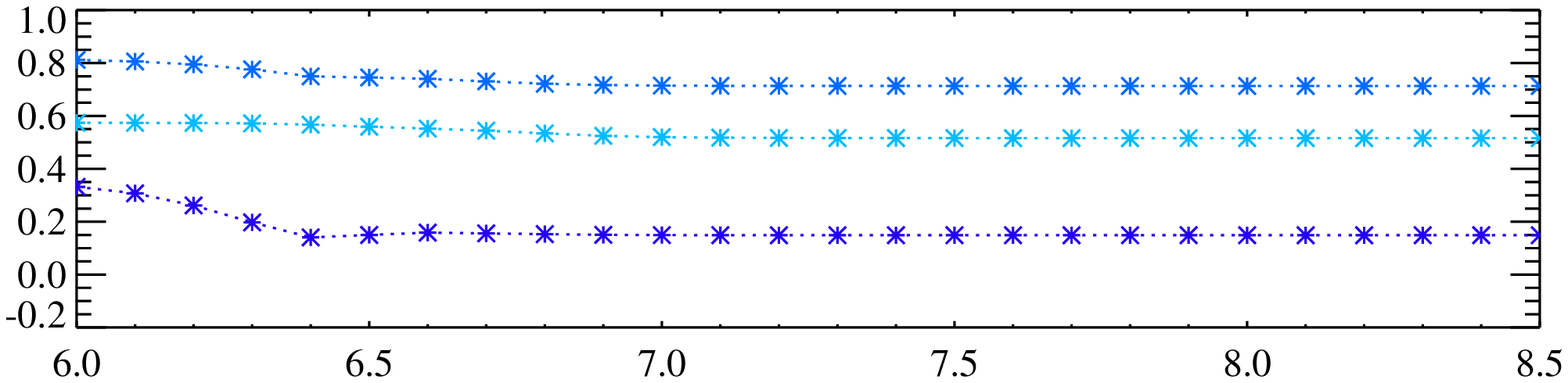}
 \includegraphics[width=\columnwidth]{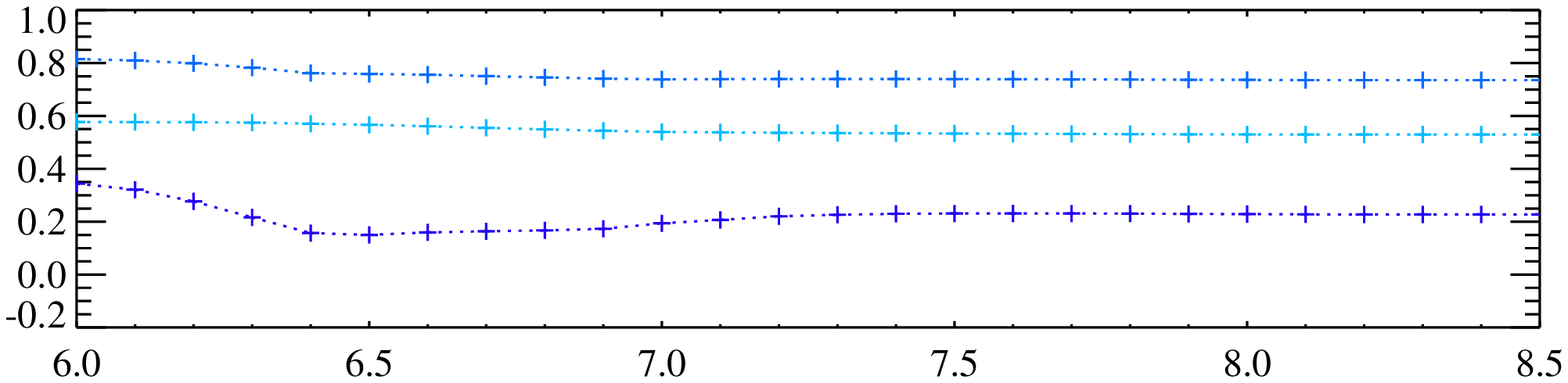}
 \caption{\label{fig:models}The distribution of  [O\,{\sc III}]/H$_\beta$ ratios predicted by BPASS models as a function of stellar population age for single star (left) and binary (right) population stellar evolution pathways. Upper panels show the ageing of an instantaneous burst. Lower panels show the near-constant ratios seen in populations with stable star formation rates. Tracks are shown at three different metallicities.}
 \end{figure*}

 \subsection{Varying \rat\ with age and metallicity}\label{sec:metal}

\begin{figure*}
 \includegraphics[width=\columnwidth]{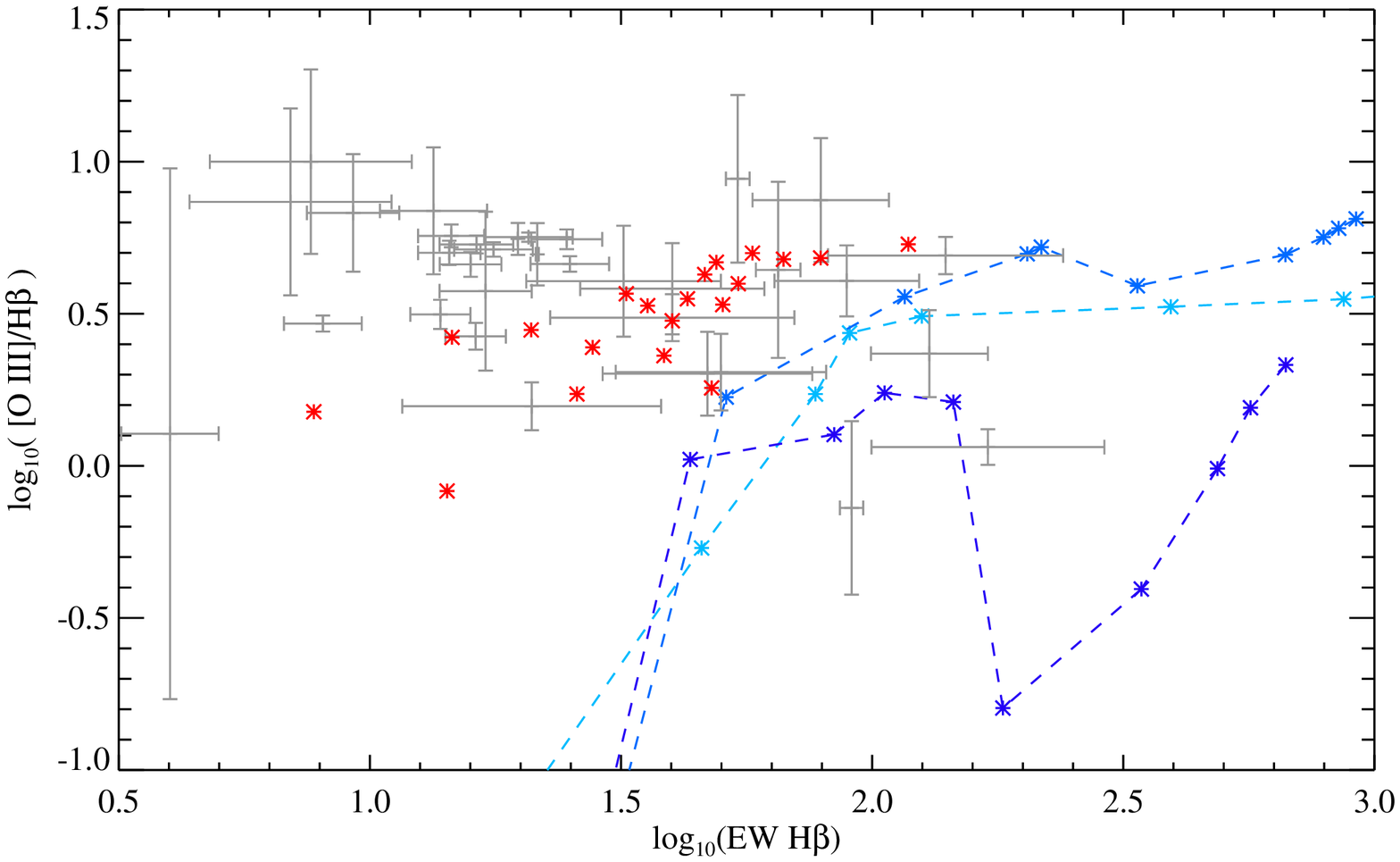}
 \includegraphics[width=\columnwidth]{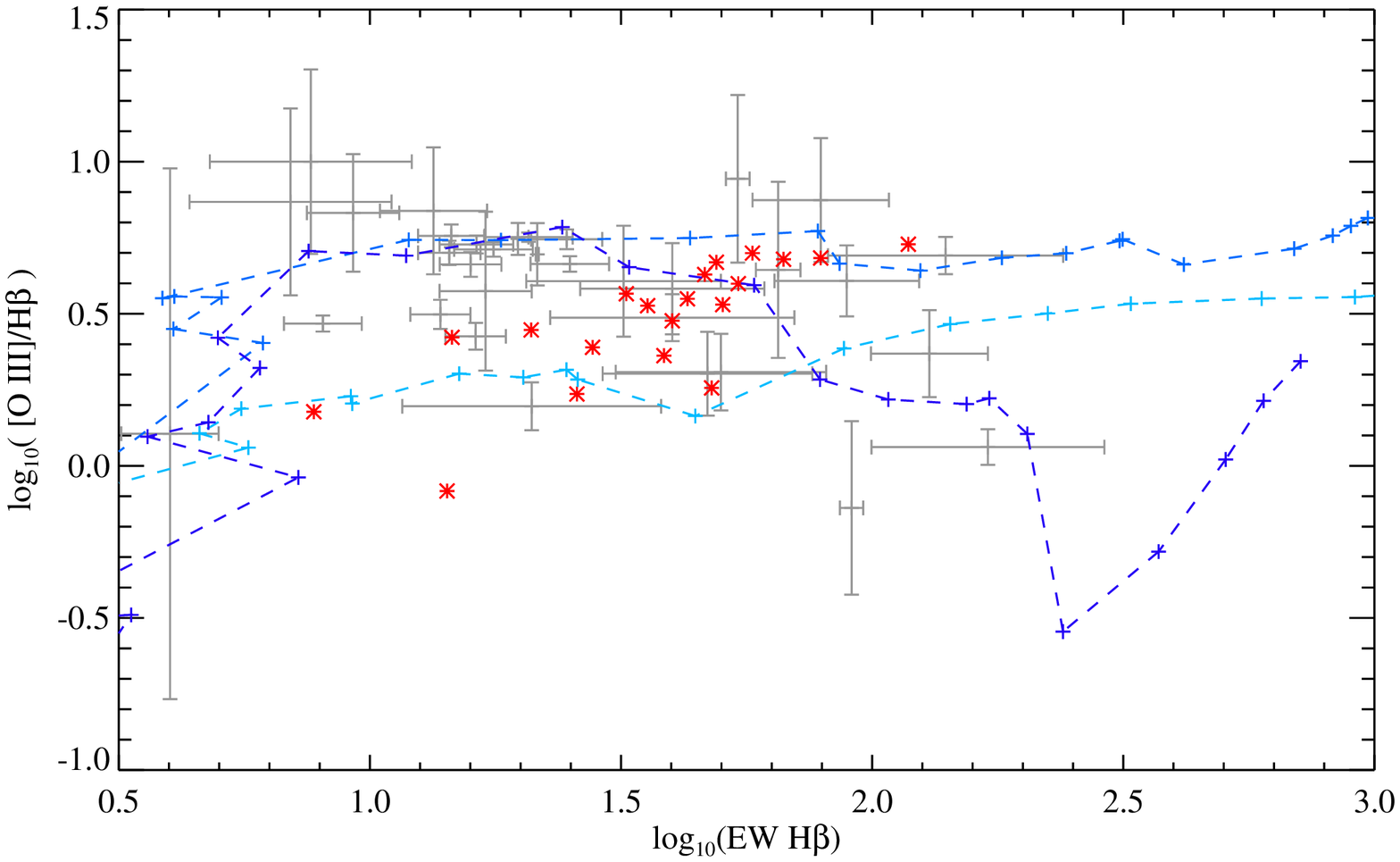}
 \includegraphics[width=\columnwidth]{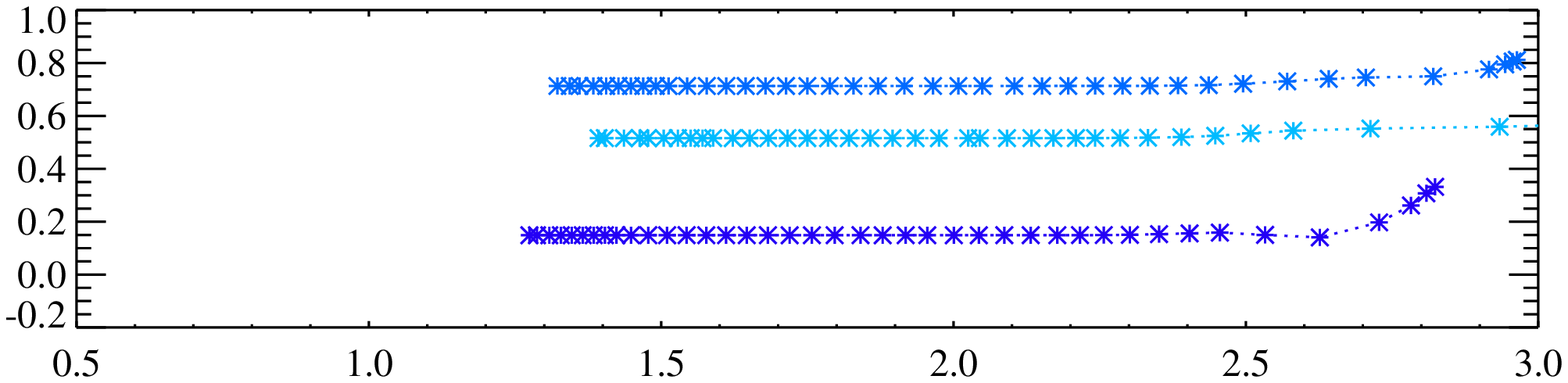}
 \includegraphics[width=\columnwidth]{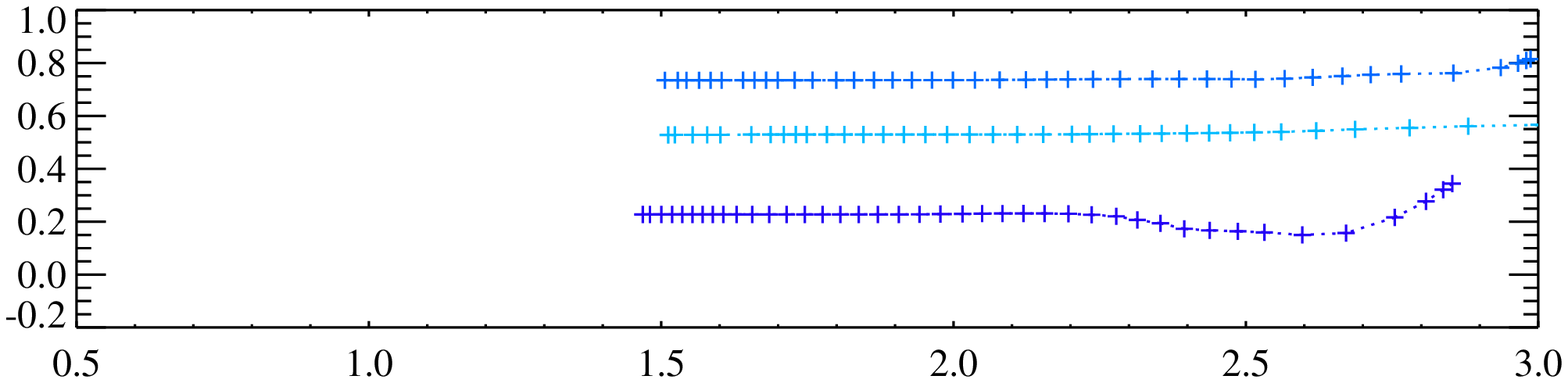}
 \caption{\label{fig:models2}The distribution of  [O\,{\sc III}]/H$_\beta$ ratios predicted by BPASS models as a function of rest-frame H\,$\beta$ recombination line equivalent width for single star (left) and binary (right) population stellar evolution pathways. Labels and linestyles are as in figure \ref{fig:models}. Age intervals of log(age)=0.1 are marked, and age increases to the left on each track in this parameter space. Pale grey points indicate high redshift data from \citet{2013ApJ...777...67S} and \citet{2014arXiv1401.5490H}. Red crosses indicate our local analogue sample, the error bars for which are smaller than the points.}
 \end{figure*}

The time evolution of the \rat\ ratio, at a fixed hydrogen gas density, is significantly affected by the introduction of binary stellar evolution pathways, as shown in figure \ref{fig:models}.  In both single and binary population models, a continuous star formation rate leads to a very stable \rat\ ratio, which does not vary significantly after the initial few Myrs, since the line flux is always dominated by the youngest stars. This ratio can reproduce the observed high redshift (and local analogue data) at sub-Solar metallicities, for either single or binary populations, but at Solar metallicity, the predicted line ratio falls well below the measured values. Continuous star formation, observed at late times, may also be consistent with the bulk of observed H\,$\beta$ emission line equivalent widths (as shown in figure \ref{fig:models2}), but struggles to reproduce some of the lower observed values.

By contrast, the aging of a rapid burst leads to line ratios that are high at early ages, as the massive stars formed in the initial burst evolve off the main sequence and eventually undergo supernovae. In single star population models, these intervals of high line ratios (\rat $> 1$) are brief, lasting no more than $\sim10$\,Myr and occuring primarily at low metallities (5-40\% of Solar). Such young stars are likely dust embedded and heavily extincted, making this an unattractive interpretation.

However, the evolution of a binary stellar population is very different. While the initial phase of high \rat\ ratios is still seen, it is extended over a much longer epoch by the formation first of lower mass Wolf-Rayet stars than are possible in a single star population, and then longer-lived, hot, helium stars. A binary stellar population will exceed a 1:1 line ratio for up to 100\,Myr after its initial formation epoch (roughly the lifetime of the minimum mass considered for binary stars in BPASS), and at much higher metallicities ($\sim$40-100\% Solar) than those seen in the single star case. At 40\% Z$_\odot$, the binary population reaches \rat\ line ratios $>3$ for  much of its first 100\,Myr, improving the probability that these ratios will be observed in conditions where a bursty, binary rich population may dominate the emission. 

{\color{black}During this interval, as figure \ref{fig:models2} shows, a binary population model also generates Balmer line luminosities and equivalent widths consistent with those seen in both the high redshift sample and the low redshift analogue population presented here. The single star instantaneous burst model, by contrast, cannot reproduce the line ratios without overproducing H\,$\beta$ line flux. Similarly, as noted above, the continuous star formation models (with or without binaries) struggle to reproduce the weakest H\,$\beta$ lines while simultaneously maintaining the strong line ratios. For both instantaneous and continuous star forming models, the lowest line equivalent widths (and luminosities) are generated at late times, as the continuum contribution from older underlying stars builds up. }

 \subsection{Varying \rat\ with gas density}\label{sec:gas}

While the emission line strengths from star forming regions are a stong function of irradiating spectrum, they also depend on the geometry and density of the emitting gas. Self-shielding and collisional excitation or de-excitation of ions can alter the transition probabilities in either very sparse or very dense gas.  As mentioned in section \ref{sec:bpass2}, the baseline BPASS model set uses a total Hydrogen gas density of $10^2$\,cm$^{-3}$, distributed in a sphere around the stellar population to model the radiative transfer and nebular emission. While this is a reasonable gas density for extragalactic star forming H\,{\sc II} regions \citep[see, for example,][]{2009A&A...507.1327H}, it is possible that a difference in the typical gas density, rather than in the dominant irradiating spectrum, could be responsible for the line properties of the high redshift starburst population and their local analogues.

\begin{figure*}
 \includegraphics[width=\columnwidth]{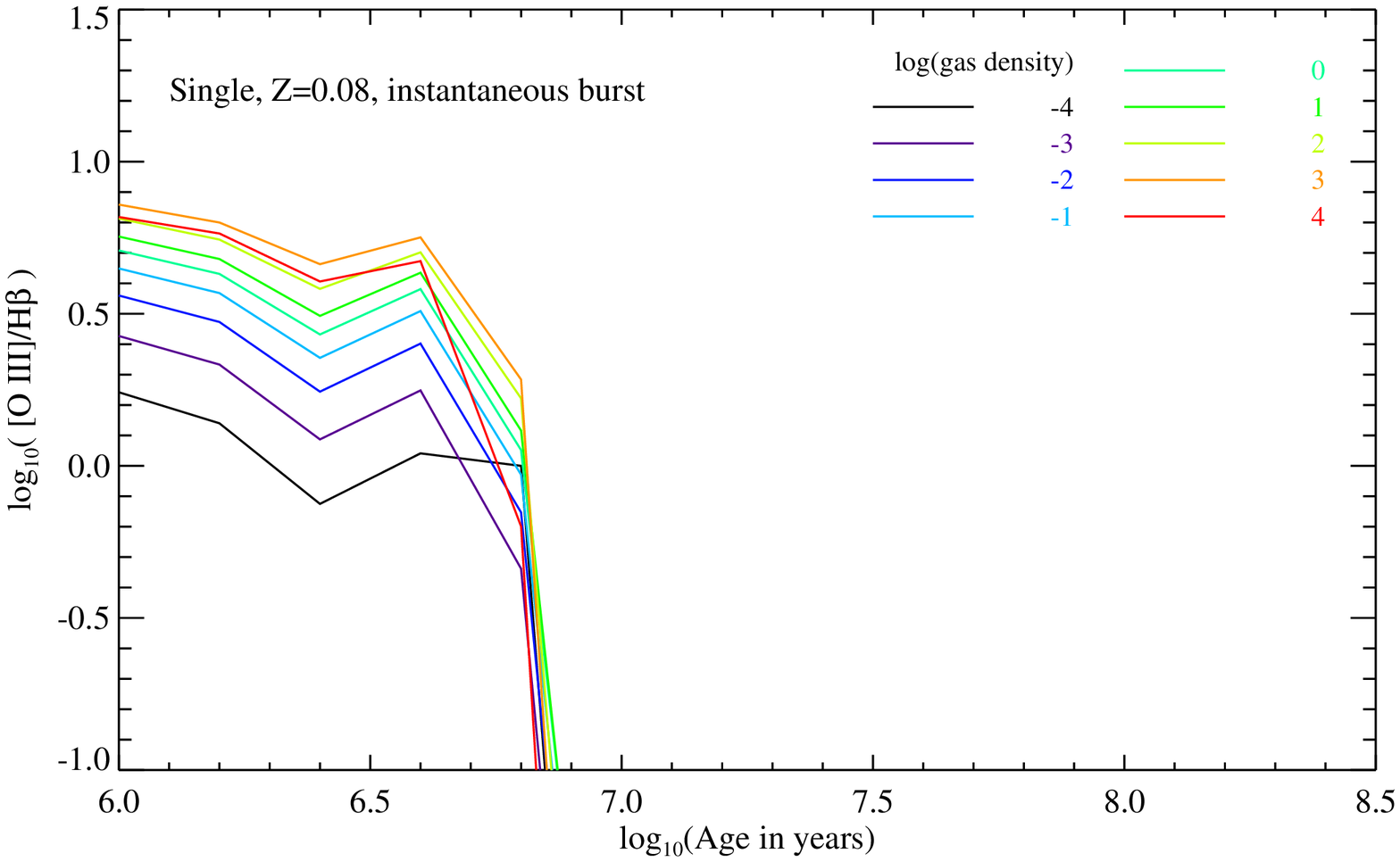}
 \includegraphics[width=\columnwidth]{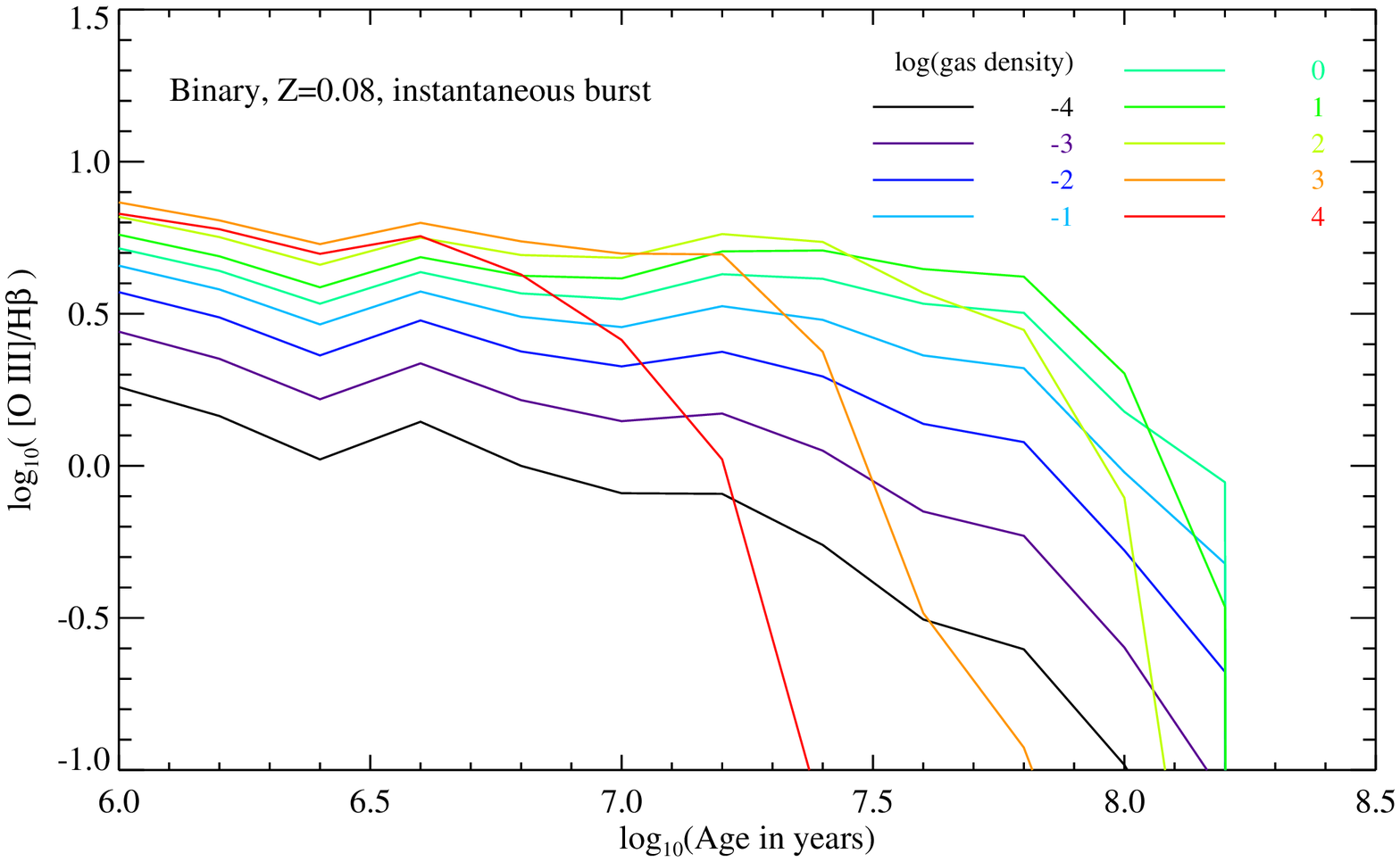}
  \caption{\label{fig:density}The distribution of  [O\,{\sc III}]/H$_\beta$ ratios predicted by BPASS models for single star (left) and binary (right) population stellar evolution pathways at $Z=0.08$ (40\% Solar), as a function of gas density. }
 \end{figure*}

In figures \ref{fig:density} we consider the effect of total hydrogen gas density on the line ratios as a function of age, for the instantaneous, Z=0.08 starburst models that best fit the data at our baseline gas density (see section \ref{sec:metal}).

The effect of changing the assumed density of the illuminated gas has a negligible effect on the lifetime of regions with high line ratios for single star populations - at all densities, the high line ratio epoch ends within 10\,Myr of the onset of star formation. The effect of gas density is, however, seen in the strength of the initial \rat\ line ratios measured. The line ratio at a given age decreases systematically with decreasing gas density, except at the highest densities considered, $10^4$\,cm$^{-3}$ - equivalent to the upper end of the HII region distribution. 

By contrast, when binary stellar populations are considered, both the strength of the ratio and the duration over which it remains elevated are functions of the assumed Hydrogen gas density. At low densities, $<$1\,cm$^{-3}$, the \rat\ ratio reached by the population increases with gas density, as does the lifetime over which it remains at the levels seen in the distant population. At higher densities, $>$1\,cm$^{-3}$, the \rat\ ratio remains very nearly constant at \rat$\sim5$, but the epoch over which this level is reached becomes shorter with increasing gas density. At a density of $10^4$\,cm$^{-3}$, the lifetime of strong line emission is comparable to that seen in the single star populations.

Comparison with figure \ref{fig:ratios} suggests that moderate gas densities, $10^{-2}-10^{4}$\,cm$^{-3}$ are required to reproduce the high line ratios seen in the distant population and their local analogues. The extended lifetime of enhanced \rat\ emission in  binary populations at gas densities $\sim1-100$\,cm$^{-3}$ and at moderate metallicities suggest that these properties may be consistent with the distant population. We note that this is not necessarily the scenario with the highest ionization parameter (which would correspond to the lowest density for a given stellar input spectrum) but is comparable to the densities seen in star forming regions.

\section{Discussion}\label{sec:discussion}

\begin{figure}
 \includegraphics[width=\columnwidth]{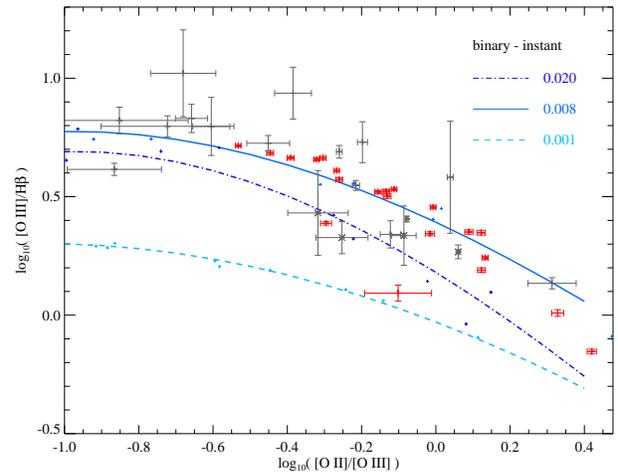}
 \caption{The [O\,{\sc II}]/[O\,{\sc III}] ratios seen in the data, and predicted by BPASS models for binary populations at ages 10$^7$-10$^{8.3}$\,yrs following an instantaneous burst. A smooth third order polynomial has been fit through the predictions at discrete model timesteps at each metallicity. Grey points are $z=2-3$ Lyman break galaxies \citep{2001ApJ...554..981P,2014ApJ...785..153M}, red points are our Lyman break analogue sample. \label{fig:oxyratio}}
 \end{figure}

\begin{figure}
 \includegraphics[width=\columnwidth]{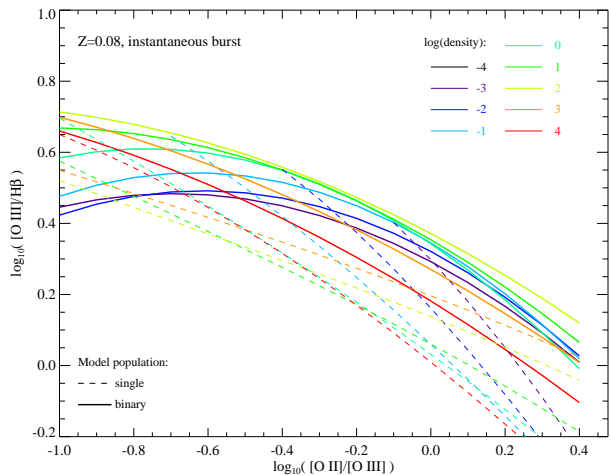}
 \caption{As figure \ref{fig:oxyratio}, but now varying assumed gas density at a fixed metallicity ($Z=0.08$). Single star models are plotted as dashed lines, fitted at ages of $10^6$-$10^{7.2}$\,yrs after an instantaneous burst. For binary populations (solid lines), gas density has only slight effect on the line ratios, such that the models occupy a narrow locus in parameter space. The single star models span a broader range, and show a more pronounced evolution in line ratios with gas density.
\label{fig:oxyratio2}}
 \end{figure}

The effect of strongly ionizing spectra (as manifest in the \rat\ ratio) has traditionally been attributed to an AGN contribution and is commonly seen in local Seyfert-type galaxies \citep{1983ApJ...264..105F}. An alternate explanation could conceivably be a very high oxygen abundance relative to hydrogen.  To rule out this later possibility, we plot both observed examples of \rat\ ratios and predictions for binary stellar populations in the BPASS models against the largely abundance-independent [O\,{\sc II}]/[O\,{\sc III}] ratio \citep[see e.g.][]{1983ApJ...264..105F} in figures \ref{fig:oxyratio} and \ref{fig:oxyratio2}. A binary star formation model at $Z=0.08$ and a gas density of 10$^{2}$\,cm$^{-3}$ (our baseline model) provides a remarkably good fit to the distribution of line ratios seen in both high and low redshift starburst galaxies, over an extended period of order 100\,Myrs. In fact, the line ratios generated by binary stellar populations of ages $<100$\,Myr shows only mild dependence on the surrounding gas density. The ratios do however show a strong dependence on binary as opposed to single star evolution pathways, with the latter typically producing lower line ratios and showing a stronger gas density dependence.

We find that high \rat\ ratios fall naturally out of a self-consistent treatment of binary evolution in an ageing starbust stellar population, without invoking AGN emission or an unusual IMF \citep{2014ApJ...785..153M}.  {\color{black}The modest star formation rates and metallicities required to create such spectra do not require fine tuning of the conditions, but are well matched to the typical properties of both $z\sim2-4$ Lyman break galaxies \citep[e.g.][]{2014A&A...563A..81D} and lower redshift UV-selected Lyman break analogues \citep[e.g.][Greis et al, in prep]{2014MNRAS.439.2474S}, while being atypical of the $z=0$ galaxy population}. Given the discrete sampling in metallicity and age of the BPASS models, and the simple prescription for the dust and gas screen, it is uncertain whether the effect of binaries is sufficient to reproduce the highest observed line ratios, \rat $\sim10$ in the $z=2-4$ sample. However, the median object is well matched to post-starburst BPASS models at $\sim$0.4\,Z$_\odot$.

\citet{2014arXiv1401.5490H} argued against a bursty starburst model for $z=2-4$ galaxies on the basis of consistency between star formation rates derived from H\,$\beta$, dominated by the youngest stellar population, and the rest-frame ultraviolet continuum. However, the time-scale for establishment of ultraviolet emission is of order $\sim10-30$\,Myr. This would be problematic for a single stellar population, but is significantly shorter than the time-scales for elevated \rat\ ratios in binary populations, which would show an established UV continuum. {\color{black}As noted above, we have studied the BPASS prediction for Balmer line luminosity and equivalent width and find it consistent with the observed high redshift data.}

So is young, bursty star formation a good model for distant galaxies with high specific star formation rates?  To some extent, any answer depends on definitions.  Degeneracies in the possible interpretation of photometric data make the `age' of a galaxy difficult to constrain. Given the ultraviolet selection of Lyman break galaxies (and analogues), they inevitably have a component of relatively young ($<1$\,Gyr) stars, but many also show evidence for an older, underlying stellar population \citep[e.g.][]{2014A&A...563A..81D,2005ApJ...626..698S,2007MNRAS.374..910E,2005MNRAS.364..443E}.  Since stellar ages are typically derived from continuum photometry, stretching into the rest-frame near-infrared, they reflect the population dominating the galaxy mass rather than the young stars (which drive nebular emission lines).

Nonetheless, SED fitting of multiwavelength data for Lyman break galaxies at $z\sim3-6$ produces typical ages of a few tens or hundreds of Myr, depending on the assumed star formation history \citep{2014A&A...563A..81D}. \citeauthor{2001ApJ...562...95S} suggest that most $z=3$ galaxies may have {\color{black}undergone a very rapid early burst on time-scales of $50-100$\,Myr, before continuing to form stars at a significantly lower rate. In such a paradigm, the evolution of that early burst may dominate the nebular continuum at ages $>300$\,Myr}. While the SEDs of LBGs at $1.4<z<2.6$ \citep{2012ApJ...754...25R} yield median stellar population ages of around 500\,Myr - somewhat older than the populations discussed in section \ref{sec:bpass} - near-infrared photometry of $z=3$ sources yields a younger median age \citep[$\sim$320\,Myr,][]{2001ApJ...562...95S}.  

A recent study at intermediate redshifts, $2.0<z<2.6$, by \citet{2014arXiv1405.5473S} has also concluded that both hard stellar radiation fields and a high ionization parameter (i.e. low gas density with relative to number of ionizing photons) are required to reproduce the distribution of line ratios in their data. As \citeauthor{2014arXiv1405.5473S} comment, our BPASS models generate a similar ionizing spectral continuum to their assumed blackbody source (with effective temperature $T_{eff}\sim42,000$\,K) and the inclusion of binary evolution pathways and stellar rotation is necessary to generate plausible ionizing spectra. We note that \citeauthor{2014arXiv1405.5473S} also identified the ``extreme green pea'' sample of \citet{2013ApJ...766...91J} as having comparable line ratios to their $z\sim2.5$ sample, a property which appears to be shared by the galaxies we have selected as analogues to $z\sim5$ star-forming galaxies.

It is likely that the hot, low metallicity star-burst with associated binary evolution presented here is not a unique explanation. As discussed in section \ref{sec:metal}, models derived for continuous star formation at low metallicity also recover high emission line ratios for a large fraction of their lifetime, although they show some tendency to overpredict the H\,$\beta$ equivalent width. Some fraction of ongoing star formation, following an initial starburst would also likely recover similar line ratios. Any stellar population synthesis code necessarily explores a limited parameter set, exploring discrete stellar metallicities, interstellar gas densities, geometries and ages. By contrast real galaxies are a composite of many star forming regions, each of different age and with different physical conditions. At best, a theoretical model can only be an approximate match. 

The high line ratios in post-starburst conditions may well be diminished by combination with other stellar populations. It is also challenging to entirely rule out a weak AGN contribution to the emission lines, although we note that X-ray stacking analyses have constrained the moderate AGN fraction in the distant population to be $<3$\% at $z\sim3$ \citep{2006MNRAS.373..217L}, inconsistent with the ubiquitous high line ratios.   Nonetheless, the ability of binary stellar population synthesis models to match the properties of distant galaxies over a reasonable, few 100\,Myr, period is encouraging, and suggests that the hard ionizing spectra of these populations may plausibly play a significant role in the evolution of the ISM (and potentially IGM) at early times.

\section{Conclusions}\label{sec:conc}

Our main conclusions can be summarised as follows:
\begin{enumerate}
\item We measure the ionization-sensitive \rat\ ratio in ultraviolet-luminous local galaxies selected as $z\sim5$ Lyman break analogues. We find they lie well above the local average ratios for their SED-derived masses, with a median \rat $=3.36^{+0.14}_{-0.04}$, similar to those seen in high redshift galaxy populations. 

\item We consider the line ratios derived from the Binary Population and Spectral Synthesis (BPASS) models. We determine that they can reproduce high \rat $\sim3$ ratios for an extended period, at ages $\sim50-300$\,Myrs, at modest (0.2-1.0\,Z$_\odot$) metallicities. They also accurately predict the behaviour of the [O\,{\sc II}]/[O\,{\sc III}] line ratio.

\item The density of the illuminated nebular gas appears to have only small effects on the predicted line ratios in binary stellar evolution models at moderate metallicities (0.4\,Z$_\odot$), and models can reproduce the data at densities seen in extragalactic star forming regions. Single stellar models are rather more sensitive to gas density, with the predicted line ratios decreasing strongly with density at a given age.

\item While continuous star formation can generate similarly high line ratios at moderate metallicities, they struggle to reproduce the measured H\,$\beta$ luminosities and equivalent widths. 

\item We conclude that including binary population effects may be important when modelling stellar populations at $<500$\,Myrs, where line ratios depend sensitively on the evolution of the most massive stars.

Ideally, the comparison of multiple emission lines, or better still modelling of the full near-ultraviolet/optical spectrum, will be required to further constrain the star formation history and properties of distant galaxies. We plan to explore those of the local analogue population further in a forthcoming paper, in the hopes of gaining further insights into possible explanations of their ionizing spectra.

\end{enumerate}

\section*{Acknowledgments}
ERS acknowledges partial funding under STFC grant ST/L000733/1.

Based in part on public data from the Sloan Digital Sky Survey DR7. Funding for the SDSS and SDSS-II has been provided by the Alfred P. Sloan Foundation, the Participating Institutions, the National Science Foundation, the U.S. Department of Energy, the National Aeronautics and Space Administration, the Japanese Monbukagakusho, the Max Planck Society, and the Higher Education Funding Council for England. The SDSS Web Site is http://www.sdss.org/. The SDSS is managed by the Astrophysical Research Consortium for the Participating Institutions. 
Calculations were performed with {\sc Cloudy}, last described by \citep{2013RMxAA..49..137F}. We also thank the anonymous referee for their input.

\bsp

\label{lastpage}

\end{document}